\documentclass[prd,aps,floats,twocolumn,showpacs]{revtex4}
\usepackage[dvips]{epsfig}
\begin{document}
\voffset 15mm
\title{Dynamics of Phase Transitions by Hysteresis Methods I}

\author{ Bernd A. Berg$^{\rm \,a,b}$, Urs M. Heller$^{\rm \,c}$,
Hildegard Meyer-Ortmanns$^{\rm \,d}$
and Alexander Velytsky$^{\rm \,a,b}$ }

\affiliation{ (E-mails: berg@hep.fsu.edu, heller@csit.fsu.edu, 
h.ortmanns@iu-bremen.de, velytski@csit.fsu.edu)\\ 
$^{\rm \,a)}$ Department of Physics, Florida State University,
  Tallahassee, FL 32306\\
$^{\rm \,b)}$ School of Computational Science and Information 
Technology, Florida State University, Tallahassee, FL 32306\\
$^{\rm \,c)}$ American Physical Society, One Research Road, 
Box 9000, Ridge, NY 11961 \\
$^{\rm \,d)}$ School of Engineering and Science, International
University Bremen, P.O. Box 750561, D-28725 Bremen, Germany
 } 

\date{Oct 24, 2003}

\begin{abstract}
In studies of the QCD deconfining phase transition or crossover 
by means of heavy ion
experiments, one ought to be concerned about non-equilibrium effects 
due to heating and cooling of the system. Motivated by this, we look 
at hysteresis methods to study the dynamics of phase transitions. Our 
systems are temperature driven through the phase transition using 
updating procedures in the Glauber universality class. Hysteresis 
calculations are presented for a number of observables, including 
the (internal) energy, properties of Fortuin-Kasteleyn clusters and 
structure functions. We test the methods for $2d$ Potts models, which 
provide a rich collection of phase transitions with a number of 
rigorously known properties. Comparing with equilibrium configurations 
we find a scenario where the dynamics of the transition leads to a
spinodal decomposition which dominates the statistical properties of 
the configurations. One may expect an enhancement of low energy 
gluon production due to spinodal decomposition of the Polyakov loops, 
if such a scenario is realized by nature.
\end{abstract}
\pacs{PACS: 05.50.+q, 11.15.Ha, 25.75.-q, 25.75.Nq}
\maketitle



\section{Introduction}

Quantum chromodynamics has well established phase transitions in 
certain limiting cases. In the limit of vanishing quark masses
it has the chiral phase transition from the phase of broken chiral 
symmetry at low temperatures to the chiral symmetric phase 
at high temperatures. In the limit of infinite quark masses
one finds the deconfinement transition from the $Z(3)$-symmetric
low temperature phase with confinement to the $Z(3)$-broken phase at
high temperatures, for a review see \cite{MO96}. For physical quark 
masses $m_u,m_d$ of the order of $10$ MeV and $m_s$ of the order of 
$150$ MeV it is suggested by lattice simulations~\cite{Br90} and 
effective models~\cite{MOS96} that neither a chiral nor a 
deconfinement transition occurs in the sense that there are 
thermodynamic singularities.

Lattice gauge theory investigations of the finite temperature phase 
transitions of QCD have, with some notable exceptions~\cite{MiOg02}, 
been limited to studies of their equilibrium properties, whereas 
in nature these transitions are governed by a temperature, or otherwise,
driven dynamics. Even when a proper phase transition does not exist, a 
question is whether one may expect observable remnants of the phase 
conversion because of off-equilibrium effects.

In the early universe the effects of the dynamics are most likely 
negligible, since the cooling process is determined by the Hubble 
expansion of the universe that is slow compared to the typical 
time scales of strong interactions, which are of the order of 
$10^{-23}$ sec. In heavy ion 
collisions this is different. A rapid heating (quench) of the 
nuclei at the ``little bang'' event is followed by a slower cooling 
process. The lifetime of the emerging system appears to be 
sufficiently long to equilibrate a phenomenological quark-gluon
plasma~\cite{AKMcL,Bj83}, although the dynamics in the time period
of the phase conversion may proceed out of equilibrium. Finite 
size corrections may play a role, because the system is not large 
compared to the typical spatial scale of strong interactions, 
i.e. 1~fm.  One should also address the question, whether the 
initial quench could lead to domains of distinct average $Z_3$ 
3-ality, with interfaces between them, which have relatively 
long relaxation times.


On an effective level (in the framework of the $O(4)$-model) 
one has studied dynamical effects on the chiral phase 
transition~\cite{BKT93,RaWi93}. Although the largest equilibrium 
correlation length (that of the pion with a mass of 
$\approx 137~MeV$) is not large compared to the intrinsic 
QCD scale (e.g., set by $\Lambda_{\rm MOM}$),
as result of a quenched cooling process one may get disoriented 
chiral condensates via spinodal decomposition. 
We are interested in the analogous question for the deconfinement 
transition~\cite{BHMV03}. One could get a disoriented condensate 
of Polyakov~\cite{Po78} loops and an associated production of 
low-momentum gluons.

Polyakov loops behave effectively like $3d$ spin 
variables~\cite{SY82,Gr83,Og84,GoOg85} and the Potts-model in three 
dimensions with $q=3$ states gives an effective description of 
the deconfinement transition (more sophisticated spin models 
are also considered~\cite{Og84}). By adding an external 
field~\cite{BU83}, one can represent the effect of finite 
quark masses. Even this simplification is not yet a suitable basis 
for a numerical investigation. To get confidence in our computational
methods, we simulated $q$-state Potts models in $2d$, for which a 
number of rigorous results~\cite{Ba73,BJ92} allow for cross checks. 
We set the external field to zero and choose $q=2,\,4,\,5$ and $10$,
corresponding to a weak second order, a strong second order, a weak 
first order and a strong first order phase transition, respectively. 
The difference between weak and strong second order transitions is
explained in section~\ref{sec_nr}. For a review of Potts models see 
Wu~\cite{Wu82}. 

We use hysteresis methods to investigate the phase transition in 
the Glauber~\cite{Gl63} dynamics. The universality class of Glauber 
dynamics, model~A in the classification of Ref.\cite{ChLu97},
contains local Monte Carlo (MC) updating schemes 
which imitate the thermal fluctuations of nature. Studying the 
computer time evolution of Glauber dynamics gives an overview of 
a scenario which allows for a variation of the speed of the phase 
transition. Notably, the notion of the Minkowskian time is lost 
in the conventional quantum field-theoretical formulation of
an equilibrium ensemble~\cite{Rothe} which is used in numerical
simulations. To study the time evolution of this field-theoretic 
ensemble, one has to find a way to re-introduce a proper dynamics.
The hope is that the thus generated configurations are typical
for the dynamical process.

Our observables are 
the internal energy, properties of Fortuin-Kasteleyn (FK) 
clusters~\cite{FK72}, and structure functions. The results from
equilibrium configurations are compared with those from configurations 
that are dynamically driven through the hysteresis cycles. In all
cases we find that the dynamics induces remarkably strong signals
for a spinodal decomposition. With increasing $q$ similar signals
become very weak for the equilibrium phase transition.

In the next section we discuss in more detail the basic concepts 
used in this paper. Our numerical investigations are reported in 
section~\ref{sec_nr}, where subsections deal with bulk properties,
FK clusters and structure functions. A brief summary and conclusions 
are given in the
final section~\ref{sec_conclude}. Article~II~\cite{BV} of this series 
will be devoted to a study of the 3d 3-state Potts model in an external 
magnetic field.

\section{Preliminaries} \label{sec_pre}

Our (computer) time-dependent Hamiltonian is 
\begin{equation} \label{H}
H(t) = - \beta(t)\,E
\end{equation}
where
\begin{equation} \label{energy}
E = - 2 \sum_{\langle \vec{r},\vec{r}'\rangle} 
\delta_{\sigma(\vec{r},t),\,\sigma(\vec{r}',t)}\ .
\end{equation}
Here the sum runs over all nearest neighbor sites $\vec{r}$ and 
$\vec{r}{\,'}$ and $\sigma$ takes the values $1,\dots\,,q$. 
In this paper we rely on symmetric lattices 
of $N=L\times L$ spins. For suitably chosen values of $\beta_{\min}$ 
and $\beta_{\max}$, we run the system at various cooling/heating rates 
in cycles from $\beta_{\min}$ to $\beta_{\max}$ and back. 
Hysteresis methods played some role in the 
early days of lattice gauge theory~\cite{CJR79}, but have apparently 
been abandoned. Possibly, the reason is that one does not learn much 
from a single hysteresis. However, averages over large numbers of 
heating and cooling cycles have to our knowledge not been analyzed 
in the literature.
By creating a large number of cycles, ensemble averages of dynamical 
configurations are obtained at selected temperatures $T=1/\beta$. For 
each temperature away from the endpoint of the cycles two distinct 
averages exist, one on the heating and the other on the cooling 
branch of the cycles. 

The spins are updated by an algorithm which is within the Glauber 
class. Examples are single- and multi-hit Metropolis, as well as 
heat-bath updating methods, where the lattice sites may be visited 
randomly or in some systematic order. At critical points (i.e. at 
second order transitions) the slowing down of such algorithms 
is governed by universal exponents. A counterexample is the 
Swendsen-Wang~\cite{SW87} algorithm, which updates entire 
FK clusters. Clearly, such an updating does not correspond to 
thermal fluctuations of nature. The purpose of the Swendsen-Wang
algorithm is to speed up the dynamics of second order phase 
transitions in computer simulations.

When driving the system through the transition, the phase conversion
may be dominated by metastable or unstable states of matter. If a 
system is brought into a metastable state, it will be unstable 
against finite, localized fluctuation. This scenario is called 
{\it nucleation}. It may allow the system to reach a metastable 
equilibrium before a large enough fluctuations occurs.
If the system is brought into an unstable state, infinitesimal, 
non-localized amplitude fluctuations lead to an immediate onset 
of the decay of the unstable state. This scenario is called {\it 
spinodal decomposition}. It may lead to long-range correlations, 
in a sense similar to those encountered in equilibrium close to 
second order phase transitions. 

The concept of nucleation as well as the spinodal were first
introduced by Gibbs as early as 1877, where the spinodal was
defined as a limit for metastability of fluid gases. But only
in the late 1950s it became apparent that a phase beyond the
spinodal decomposes by diffusional clustering mechanism quite
different from the nucleation and growth mechanism encountered for
metastable states. In his classic review~\cite{Ca68} Cahn includes 
an account of the historical development. The modern theory uses 
effective diffusional differential equations (originally an idea
of Hillert~\cite{Hi61}) to distinguish 
dynamical universality classes, see Ref.\cite{ChLu97,La92,Gu83}
for reviews. A sharp distinction between infinitesimal (spinodal)
and finite (nucleation) fluctuations is, strictly speaking, a 
mean field concept. In real systems, where fluctuations are 
important, the boundary separating nucleation from spinodal
decomposition is not perfectly sharp.

The numerical investigations, we are aware off, investigate the
spinodal versus the nucleation scenario after a quench, which
may either lead into the metastable region (nucleation) or beyond
it (spinodal decomposition). See Miller and Ogilvie~\cite{MiOg02} 
in the context of lattice gauge theory. Our hysteresis approach
differs in this respect. The continued change of the external
temperature prevents the system from ever reaching equilibrium,
but implies on the other hand a smoother dynamics, because the
temperature changes only in small steps. Under laboratory 
conditions there is never a perfect quench and in some situations
our hysteresis approach may allow to model laboratory condition
more realistically than a quench.
We measured many observables in each hysteresis cycle. In this 
paper we report selected results for the energy, FK 
clusters and structure functions. In more detail the data will
be analyzed and presented in Ref.\cite{Vel}.

We measure FK clusters instead of geometrical clusters, because 
their statistical definition accounts for the fact that neighboring 
spins may not only be aligned by the spontaneous magnetization 
but also by random fluctuations. It is only then that the 
Kertesz~\cite{Ke89} line of percolation coincides with the 
phase transition, see Ref.\cite{StAh94} for a review of this and
related topics. In contrast to the stochastic definition,
the geometric definition connects aligned spins with certainty 
and leads to an overcounting of ordered clusters. While the FK works 
well for Potts models, a generalization to gauge theories is not 
known. This is closely related to the fact that a cluster updating 
algorithm is not known for gauge theories.

We are interested in the effects of dynamic heating and cooling 
on the cluster structure, in particular in the question, whether 
one may still find observable signals, even when their is no longer 
a transition in the strict thermodynamic sense. There are 
similarities and differences to the program of Satz~\cite{Sa01}. 
Satz focuses on geometric properties of FK clusters and would like 
to extract from their equilibrium distribution signals for the 
phase conversion when there is no proper phase transition. We 
are trying to find signals for the phase conversion due to 
the deviations from equilibrium. For nucleation one expects 
compact clusters, due to the non-zero interfacial tension between 
the ordered and the disordered phase. 
For spinodal decomposition clusters of each of the ordered states 
will grow unrestricted by such an interfacial tension, building 
domain walls between the distinct ordered states. For nucleation 
we expect the maximum cluster surface to grow to a size $c\,L^{d-1}$
with $c\approx 2$ for strong first oder transitions ($c=2$ for the 
smallest surface of a cluster which percolates). For spinodal 
decomposition we expect considerably larger values, comparable to 
the largest values one finds on equilibrium configurations in the 
neighborhood of a second order phase transition.

In our simulations we record the following cluster observables:
their number, the mean volume, the maximum volume, the mean 
surface area, the maximum surface area, the gyration radius and 
the percolation probability. The volume of a cluster is simply the 
number of spins it contains.  The cluster surface is defined on 
the links of the dual lattice, which corresponds to the $d-1$ 
dimensional hyperspace of the original lattice. 
The percolation probability $p$ is the probability to find at least
one cluster that percolates. For our periodic lattices this means
that the cluster connects to itself through the boundary conditions, 
in any one of the two directions.


We analyze the structure function in momentum space for signals
of spinodal decomposition. Let $m_{q_0}=\langle 
\delta_{\sigma(\vec{r},t),q_0}\rangle$ denote the magnetization in
direction $q_0\in\{1,\dots,q\}$. By introducing a Potts spin
$S_{q_0}(\vec{r},t)=\delta_{\sigma(\vec{r},t),q_0}$
we can write the correlation function 
\begin{equation}
g(\vec{r},\vec{r}^\prime,t)
=\langle\delta_{\sigma(\vec{r},t),\sigma(\vec{r}^\prime,t)}\rangle-
\sum_{q_0} m_{q_0}^2,
\end{equation}
in the familiar form
\begin{equation}
g(\vec{r},\vec{r}^\prime,t)=\sum_{q_0=0}^{q-1} \left\langle 
 S_{q_0}(\vec{r},t) S_{q_0}(\vec{r}^\prime,t)\right\rangle -
\sum_{q_0=0}^{q-1} \left\langle S_{q_0}\right\rangle^2\ .
\end{equation}
The structure factor (function) is the Fourier transform of the 
correlation function
\begin{equation}
S(\vec{k},t) = \frac1{N_s}\sum_{\vec{R}}g(|\vec{R}|,t)\exp[i\vec{k}\vec{R}]
\end{equation}
where $\vec{R}=\vec{r}-\vec{r}^\prime$. Some straightforward algebra
transforms this into
\begin{eqnarray} \nonumber
S(\vec{k},t) &=& \frac1{N_s^2}\sum_{q_0=0}^{q-1}
\left\langle\left|\sum_{\vec{r}}\delta_{\sigma(\vec{r},t),q_0}
\exp[i\vec{k}\vec{r}]\right|^2\right\rangle\\ \label{sf}
 &-& \delta_{\vec{k},0} \sum_{q_0} m_{q_0}^2\ .
\end{eqnarray}
This is simply the time-dependent version of the equilibrium structure
factor. In condensed matter experiments the magnitude of the structure
function is directly observable in X-ray, neutron and light scattering
experiments, compare, e.g., Ref.~\cite{St71}. Unfortunately, it appears
to us that direct measurements in high energy experiments are 
unrealistic. In our simulations we expect pronounced peaks (similar
as for equilibrium configuration near second order phase transitions)
for $S(\vec{k},t)$ in the case of a phase conversion by spinodal 
decomposition and no such signals in the case of a conversion by
nucleation and growth.


\section{Numerical Results} \label{sec_nr}

The data presented in this paper rely on systematic updating for which 
the Potts spins are updated in sequential order, each spin once during 
one sweep. We did a number of cross-checks using random updating for
which the spins are updated in random order, in the average each spin 
once during one sweep. Besides a slowing down of the dynamics by
a factor of about 0.6 for random updating, we observed no noticeable 
changes of the results checked.

The temperature $\beta = 1/T$ is changed by $\pm\triangle\beta$ after 
every sweep (we experimented also with temperature changes after each
spin update and found no differences within our statistical errors). 
Our stepsize $\triangle\beta$ is proportional to the inverse volume 
of the system
\begin{equation} \label{delta_beta}
\triangle\beta = {2(\beta_{\max}-\beta_{\min})\over n_{\beta}\,L^2}
\end{equation}
where $\beta_{\min}$ and $\beta_{\max}$ define the terminal 
temperatures and the integer $n_{\beta}=1,2,\dots $ is varied.
Equilibrium configurations are recovered in the limit 
$n_{\beta}\to\infty$ ($\triangle\beta=0$). In nature the 
fluctuations per spin per time unit (here the unit of one MC sweep) 
set the scale for the dynamics. Our choice of $\triangle\beta$ is 
motivated by our interest in the question whether a dynamics, which
slows down with volume size may still dominate the nature of the 
transition.  Relying on the heat-bath method, each of our systems 
is driven through at least 640 cycles, each starting from an 
equilibrated, disordered configuration.  Error bars are calculated with
respect to 32 jackknife bins. In an exploratory simulation~\cite{VBH02} 
of $2d$ Potts models the Metropolis algorithm was employed, but it 
turns out that the heat-bath method saves CPU time.

In this article we simulate the $2d$ $q$-state Potts model for $q=2$, 
4, 5 and 10. This allows us to compare the influence of the Glauber
dynamics for a weak second order, a strong second order, a weak first 
order and a strong first order phase transition. Our terminology 
``strong second order phase transition'' may need some explanation.
For a finite system of volume $L^d$ the partition functions is a
polynomial in $u=\exp(-\beta)$ that takes positive values on the
real axis. For first and second order phase transitions the imaginary
part of the partition function's zero closest to the real axis scales
like $u_y^0\sim L^{-1/\nu}$, where $\nu=1/d$ for a first order 
transition and $1/d<\nu\le 2/d$ for a second order transition. The 
fluctuations of the energy are governed by the exponent $\alpha$ of 
the specific heat for which we assume the hyperscaling 
relation~\cite{Fi74} $\alpha=2-d\nu$. Therefore, $\alpha=1$ 
for first order transitions and $0\le\alpha<1$ for second order 
transitions. To determine the implications for the finite size 
scaling of the energy fluctuations, we use the link expectation 
value of the energy
\begin{equation} \label{e_l}
e_l = e_l(\beta) = \langle \delta_{\sigma(\vec{r},t),
\sigma(\vec{r}',t)} \rangle = - \langle E\rangle / (2d\,L^d)\ , 
\end{equation}
where $\vec{r}$ and $\vec{r}^{\,'}$ are nearest neighbor sites. 
The values of $e_l$ are conveniently located in the range 
$0\le e_l\le 1$ with $e_l(0)=1/q$ and $e_l(\infty)=1$. To leading 
order in $L$, finite size scaling theory predicts the fluctuation 
of $e_l$ to scale like
\begin{equation} \label{sigma_e_l}
\langle (e_l)^2\rangle-\langle e_l\rangle^2 \sim L^{\alpha/\nu -d}
\end{equation}
for $\beta$ at the transition point $\beta_c=1/T_c$. For first 
order phase transitions $\alpha=1$ holds and the left-hand-side of
equation~(\ref{sigma_e_l}) approaches a finite value, proportional 
to the square of the latent heat $\triangle e_l$. For second order 
phase transitions the left-hand-side scales to zero. In this sense 
a weak second order transition is one with $\alpha$ close to zero 
or $\alpha=0$ and a cusp or logarithmic singularity, while a 
strong second order transition has $\alpha$ close to one. First
order transitions are weak when $\triangle e_l \ll 1$ holds and 
strong when $\triangle e_l$ becomes of order one, say from 
$\triangle e_l>0.1$ on. For our choices
of $q$ the analytical values~\cite{Ba73,Wu82} of $\beta_c$, $\alpha$ 
and $\triangle e_l$ are compiled in table~\ref{tab_potts}. Our values 
of $\beta_{\min}$ and $\beta_{\max}$ for equation~(\ref{delta_beta})
and a numerical result, $\overline{\triangle e_l}$, as explained in
the following subsection, are also given in this table.

\begin{table}[ht]
\caption{ The (infinite volume) phase transition temperatures 
$\beta_c=1/T_c$, the 
specific heat exponent $\alpha$ and the latent heats of selected 
$q$-state Potts models in two dimensions. For the latent heats the 
negative energy per link $\triangle e_l$ is given and 
$\overline{\triangle e_l}$ is an estimate from hysteresis cycles. 
\label{tab_potts} }
\medskip
\centering
\begin{tabular}{||c|c|c|c|c|c|c|c||}   \hline
 $q$& $\beta_c$ & $\alpha$ &
                 $\triangle e_l$&
              $\beta_{\min}$ & $\beta_{\max}$ &
                               $\overline{\triangle e_l}$\\ \hline
  2 & 0.440687 &  0  & 0        & 0.2 & 1.0 & 0.0153 (07)\\ \hline
  4 & 0.549306 & 2/3 & 0        & 0.2 & 1.0 & 0.0907 (11)\\ \hline
  5 & 0.587179 &  1  & 0.031072 & 0.4 & 1.2 & 0.1402 (12)\\ \hline
 10 & 0.713031 &  1  & 0.348025 & 0.4 & 1.2 & 0.3482 (16)\\ \hline
\end{tabular} \end{table} 

In steps of 20 our lattice sizes range from $L=20$ to $L=100$. For
the smaller systems all hysteresis runs are done on a single PC,
while for the larger lattices up to 32 PCs are used, dividing our
entire run in 32 bins of at least 20 hysteresis loops each. In
each case a short equilibrium run of $q\,L^2$ sweeps was initially 
performed at $\beta_{\min}$, where the systems equilibrate easily, 
because they are highly disordered. For comparison with equilibrium 
configurations we performed multicanonical~\cite{BN92,BNB94} (see the
next subsection) as well as conventional, canonical simulations. The 
reason for the conventional canonical equilibrium simulations is that 
one needs to know the temperature to generate FK clusters. They
were performed at many temperatures and in each case 640 measurements
were taken after at least $20q\,L^2$ sweeps for reaching equilibrium.

\subsection{Internal Energy} \label{subsec_bulk}

\begin{figure}[-t] \begin{center}
\epsfig{figure=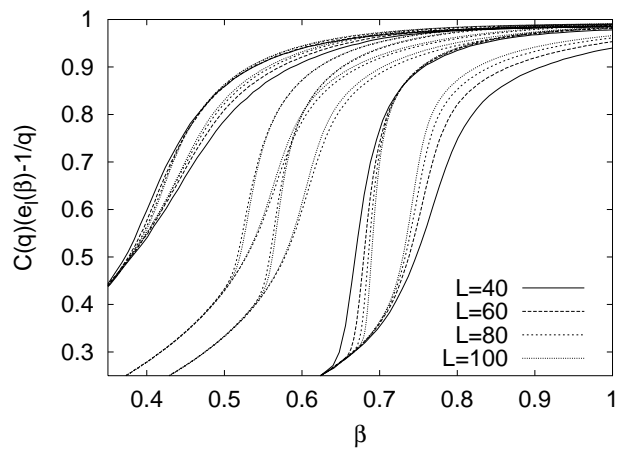,width=\columnwidth} \vspace{-1mm}
\caption{Energy~(\ref{e_l}) hysteresis curves for $n_{\beta}=1$. From 
left to right: $q=2$, 4, 5, and 10. } \label{fig_hys_e}
\end{center} \vspace{-3mm} \end{figure}

\begin{figure}[-t] \begin{center}
\epsfig{figure=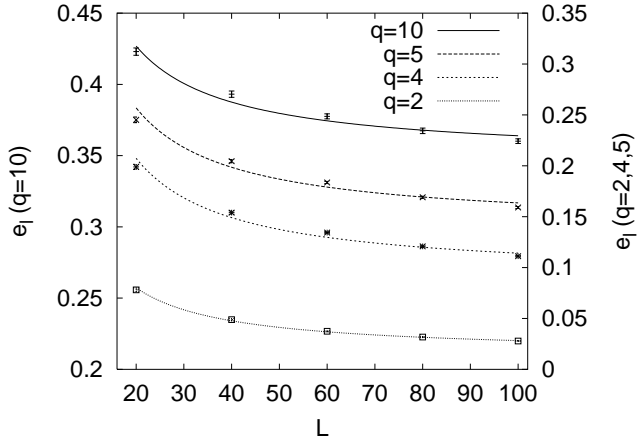,width=\columnwidth} \vspace{-1mm}
\caption{Latent heat estimates from $n_{\beta}=1$ hysteresis curves.} 
\label{fig_dele}
\end{center} \vspace{-3mm} \end{figure}

For a first order phase transition, the slowing down of the canonical
equilibrium Markov process is exponential in computer time, 
$\sim \exp [2\,f_sL^{d-1}]$, where $f_s$ is the interfacial tension 
(see~\cite{BJ92} for the analytical values). 
In this case we expect an energy hysteresis to survive in the limit 
$L\to\infty$ and $\triangle \beta (L;n_{\beta})\to 0$ for any fixed 
value of $n_{\beta}$ in equation~(\ref{delta_beta}).
The shape of the hysteresis can then be used to define finite volumes 
estimators of physical variables, such as the transition temperature 
and the latent heat. The infinite volume limits of these estimators
are supposed to be independent of any fixed choice of $n_{\beta}$.

For the second order phase transitions of the $q=2$ and $q=4$ models 
the analysis is more subtle. The Markov process slows only down like 
$L^z$ with $z\approx 2$~\cite{BlNi00}. Therefore, one still expects a 
hysteresis in the limit $L\to\infty$ and $n_{\beta}$ fixed, only the 
opening has no longer the interpretation of a finite volume estimator 
of the equilibrium latent heat. A finite size scaling analysis of the 
hysteresis as function of $n_{\beta}(L)$ should allow to identify 
second order transitions. This analysis is not pursued here.

For $n_{\beta}=1$ and selected lattice sizes we show in 
figure~\ref{fig_hys_e} our energy~(\ref{e_l}) hysteresis data. The
ordinate is scaled to 
\begin{equation}
C(q)\,\left( e_l(\beta)-{1\over q} \right)~~
{\rm with}~~C(q)={q\over q-1}
\end{equation}
so that, independently of $q$, the range $[0,1]$ gets covered when
$\beta$ is varied from $0\to\infty$.
If one wants to compare the present heat bath with Metropolis 
results~\cite{VBH02}, the better efficiency of the heat bath algorithm 
is such that $n_{\beta}^{\rm Metropolis}\approx 
q\,n_{\beta}^{\rm heat bath}$ ought to be used. From left to right in 
figure~\ref{fig_hys_e} hysteresis loops for the cases $q=2$, 4, 5 and~10
are visible. For clarity of the figure we have omitted error bars and
for $q=4$ and~5 also the $L=40$ and $60$ lattices. Notable is that the
hysteresis curves for the $q=4$ strong second order transition and
the $q=5$ weak first order transition are quite similar. From $q=10$
to $q=2$ there is a gradual, not an abrupt, deformation of the shape
of the hysteresis.

To analyze the physical content of the hysteresis curves of 
figure~\ref{fig_hys_e} in more detail, we define the finite volume 
estimators of the inverse transition temperature $\beta_c(L)$ and of 
the latent heat $\triangle e_l(L)$ by their values at the maximum 
opening of the corresponding hysteresis curve. Figure~\ref{fig_dele} 
shows the thus obtained estimates $\triangle e_l(L)$ together with 
fits of the form 
\begin{equation} \label{del_efits}
\triangle e_l(L) = \overline{\triangle e_l}+{a_1\over L}
\end{equation}
where $a_1$ is a constant. For $q=10$ the left-hand-side ordinate
applies and for the other $q$-values the right-hand-side ordinate.
Because of the distinct scales the difference between the $q=10$
and the $q=5$ estimators is large, while the general behavior of
the fitting curve appears to be quite similar for all $q$-values.
The obtained infinite volume estimates $\overline{\triangle e_l}$ 
are given in table~\ref{tab_potts}. For $q=10$ the estimate 
is in excellent agreement with the analytical result, but this 
is not at all the case for the other 
$q$-values. Instead, the $q=2$, 4 and~5 estimates overshoot the
equilibrium values considerably. This does not come as a surprise, 
because we already noted that, in the infinite volume limit and for 
fixed $n_{\beta}$, a finite opening of the hysteresis survives even 
for the second order phase transitions. Obviously, the opening has 
no longer the interpretation of an estimator of the equilibrium 
latent heat. Instead, the phenomenon illustrates that the dynamics 
tends to wash out differences of the equilibrium properties of the 
transitions.

Performing a similar analysis for $\beta_c(L)$ and comparing the
infinite volume estimates with the analytical results, we get
accuracies of about $\pm$1\% for all $q$. So, we find no problem in 
locating the equilibrium transition temperature from the information 
of the dynamics. The accuracy of these dynamical estimates is not 
competitive with the best equilibrium methods. E.g., fitting 
the pseudocritical $\beta$-values of the multicanonical 10-state 
Potts model simulation~\cite{BN92} self-consistently to the form 
$\beta_c(L)=\beta_c+c/L^2$ gives $\beta_c = 0.713032\,(16)$ (using 
our energy convention~(\ref{energy})). 
The purpose of our present study is not to calculate high precision
estimates of equilibrium quantities, but to investigate the 
deviations from equilibrium due to the imposed dynamics.
To understand the dynamics of our finite 
volume transitions in more detail, we analyze in the next two 
subsections the behavior of FK clusters and structure functions on 
our configurations. 

In a last remark about the hysteresis curves of the internal energy,
we like to mention that we have also generated equilibrium data for 
all cases using the multicanonical method. As expected, the thus 
obtained $e_l(\beta)$ functions fall inside the hysteresis curves 
of figure~\ref{fig_hys_e}. Some more details are given in
Ref.\cite{VBH02}.

\subsection{Cluster Properties} \label{subsec_cluster}

\begin{figure}[ht] \vspace{-2mm} \begin{center}
\epsfig{figure=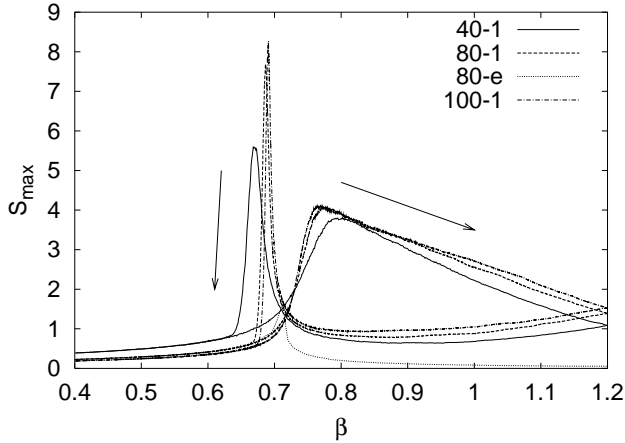,width=\columnwidth} \vspace{-1mm}
\caption{The largest cluster surface for the 10-state Potts 
model on various lattice sizes as indicated in the figure (the
extensions are the value of $n_{\beta}$ and e for equilibrium). } 
\label{fig_q10_sm} \end{center} \vspace{-3mm} \end{figure}

\begin{figure}[ht] \vspace{-2mm} \begin{center}
\epsfig{figure=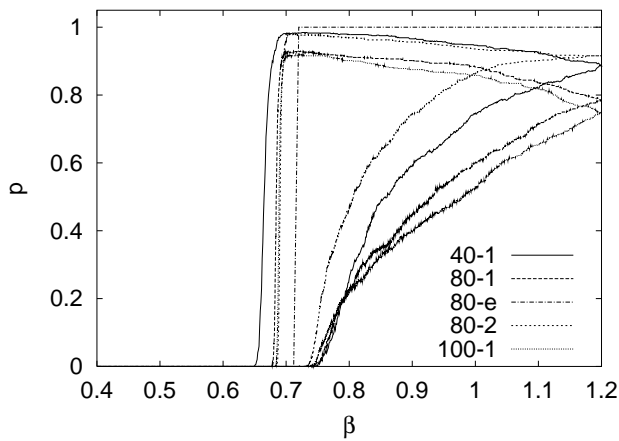,width=\columnwidth} \vspace{-1mm}
\caption{The probability $p$ of having a percolating cluster for 
the 10-state Potts model on various lattices. } 
\label{fig_q10_prc} \end{center} \vspace{-3mm} \end{figure}

\begin{figure}[ht] \vspace{-2mm} \begin{center}
\epsfig{figure=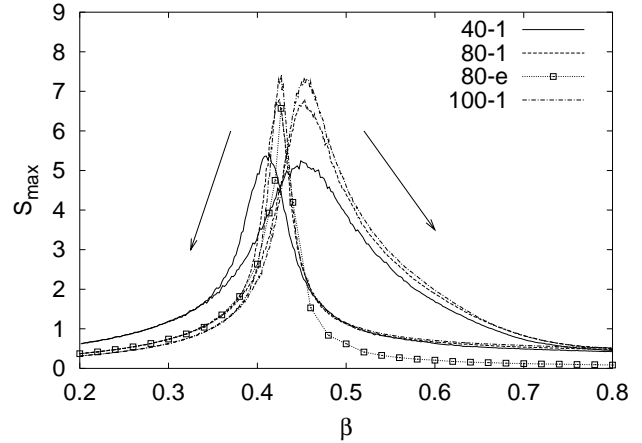,width=\columnwidth} \vspace{-1mm}
\caption{The largest cluster surface for the 2-state Potts 
model on various lattice sizes as indicated in the figure. } 
\label{fig_q02_sm} \end{center} \vspace{-3mm} \end{figure}

\begin{figure}[ht] \vspace{-2mm} \begin{center}
\epsfig{figure=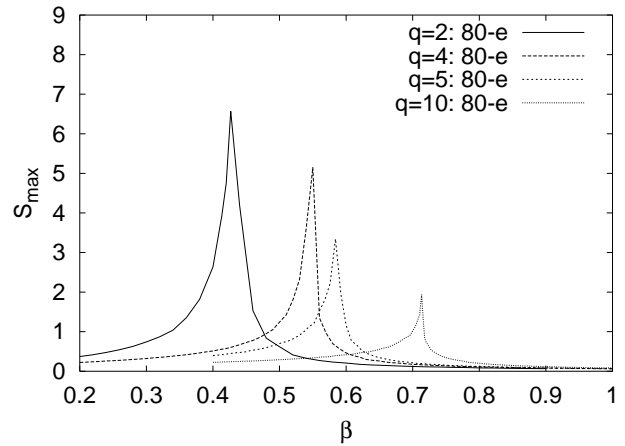,width=\columnwidth} \vspace{-1mm}
\caption{The largest cluster surface for equilibrium simulations.} 
\label{fig_eq_sm} \end{center} \vspace{-3mm} \end{figure}

We limit our presentation to a few of the cluster observables we
measure (more details will be given in~\cite{Vel}). The largest
cluster surface turns out to be interesting, because it exhibits 
pronounced peaks in the transition region. We use the normalization 
\begin{equation}
 S_c = {\# {\rm \ surface\ links\ of\ cluster\ } c \over L^{d-1} } 
\end{equation}
for our cluster surfaces. A link which connects a site of the
cluster with the site of another adjacent cluster is defined to 
be a surface link. The surface links can be mapped on the 
$(d-1)$-dimensional hypercubes which enclose the cluster on 
the dual lattice. The largest surface is simply defined as
\begin{equation}
 S_{\max} = \max \{ S_c \} 
\end{equation}
where the maximum is taken over all clusters $c$ of the configuration
at hand.

For the 10-state model results for the largest cluster surface of the
$n_{\beta}=1$ hysteresis cycle are shown in figure~\ref{fig_q10_sm}. 
The arrows indicate the flow of the hysteresis cycles. During the 
heating and cooling parts of the cycles, the surface areas peak 
at distinct values, $\beta =\beta_{\rm peak}^{\pm}$.  
This is striking evidence that the geometry of the FK 
clusters is distinct during cooling and heating. Due to our use of
stochastic (in contrast to geometrical) clusters, the equilibrium
transition temperature value is pinched between the temperatures
at which the two peaks are located.

It can be understood that the peaks of $S_{\max}(\beta)$ are related
to percolation. For the $\beta\to\beta_{\max}$ half-cycle the picture
is that the cluster with the largest surface percolates due to the 
periodic boundary conditions. Until the cluster percolates, its surface 
area increases, while it is decreasing after percolation (as only small 
islands of the false phase remain eventually). Relying on the same data 
as for figure~\ref{fig_q10_sm}, we show in figure~\ref{fig_q10_prc} the 
percolation probability $p$. It is seen that the temperatures of the 
$S_{\max}$ peaks correspond approximately to the steepest 
increase/decrease of the percolation probabilities.

Another observation from figure~\ref{fig_q10_sm} is that for the 
half-cycle $\beta\to\beta_{\min}$ the peaks of $S_{\max}$ are even 
more pronounced than for $\beta\to\beta_{\max}$. This is in accordance 
with a very rapid fall-off of the percolation probability for the
$\beta\to\beta_{\min}$ half-cycle. Our interpretation is that the 
response to the temperature change is more rapid 
when the system enters the disordered phase than when it enters the
ordered phase. Such a change in relaxation scales may be expected for
a strong first order transition (because both phases are separated by 
a gap in the energy and not continuously related), while one would 
expect that the 
response times under heating and cooling are similar for a weak
second order phase transition.  Indeed, figure~\ref{fig_q02_sm} shows
that the two peaks are of almost equal height for the 2-state Potts 
(Ising) model. The $S_{\max}$ results for the $q=4$ and $q=5$ models 
(no figures shown) are in-between the two scenarios, but certainly 
closer to $q=10$ than to $q=2$. The difference between $q=4$ and $q=5$ 
is minor. 

Also shown in figures~\ref{fig_q10_sm} and~\ref{fig_q02_sm} are results
for $S_{\max}(\beta)$ from equilibrium simulations on $L=80$ lattices. 
They are barely visible, because they are to a large extent covered by 
the curves of the $\beta\to\beta_{\min}$ half-cycle. Therefore, we plot 
the equilibrium curves for all our $q$-values separately in 
figure~\ref{fig_eq_sm}. The peaks show a marked increase from $q=10$
(right) to $q=2$ (left).
For first order phase transitions the interface tension implies that 
the free energy increases with the cluster surfaces. The stronger the 
first order transition is, the more the system tries to minimize 
interfaces. For a second order phase transition there is no 
(disorder-order) free energy penalty when the phases mix and the 
cluster surfaces become fluffy. This is quite similar to the
distinct behavior of cluster surfaces under nucleation versus 
spinodal decomposition.  The suggestion 
from figures~\ref{fig_q10_sm} and~\ref{fig_q02_sm} is then that 
the dynamics changes the transition scenario to spinodal for all 
our $q$-values. In these cases the heights of the peaks are
quite similar to those which we find for the equilibrium peaks
of the $q=2$ and $q=4$ second order transitions.

\begin{figure}[ht] \vspace{-2mm} \begin{center}
\epsfig{figure=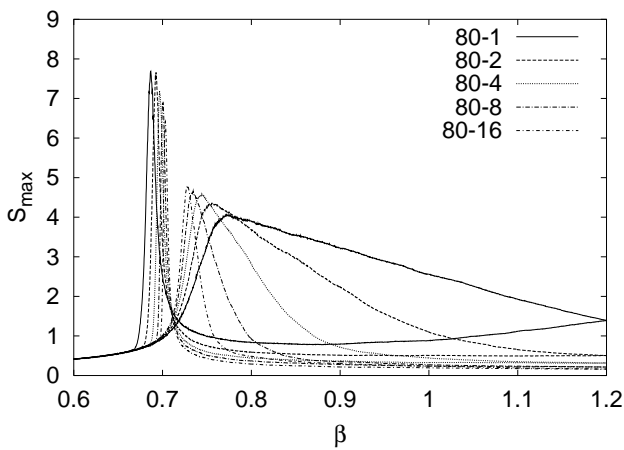,width=\columnwidth} \vspace{-1mm}
\caption{The largest cluster surface for the 10-state Potts model 
on $80\times 80$ lattices for the $n_{\beta}$ values indicated by 
the extensions to the lattice size. } 
\label{fig_q10_smV} \end{center} \vspace{-3mm} \end{figure}

\begin{figure}[ht] \vspace{-2mm} \begin{center}
\epsfig{figure=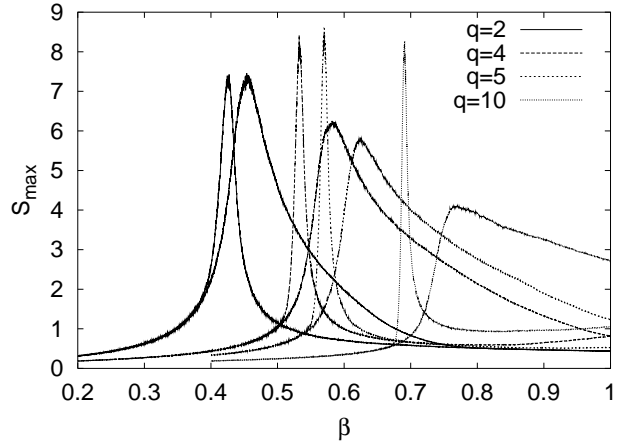,width=\columnwidth} \vspace{-1mm}
\caption{The largest cluster surface from $n_{\beta}=1$ dynamical 
simulations on $100\times 100$ lattices. Compare with the 
equilibrium results of figure~\ref{fig_eq_sm}.} 
\label{fig_1V_sm} \end{center} \vspace{-3mm} \end{figure}

The question emerges, how fast is the equilibrium scenario approached
when the speed of the dynamics slows down? In figure~\ref{fig_q10_smV}
we plot for $n_{\beta}=1$, 2, 4, 8, and~16 our $S_{\max}(\beta)$ results 
of the 10-state model on an $80\times 80$ lattice. For increasing 
$n_{\beta}$ we observe a slight decrease of the $\beta\to\beta_{\min}$
peaks, while the $\beta\to\beta_{\max}$ peaks increase. Although the
peaks of the cooling and heating half-cycles approach one another
in this way, each process is still far away from equilibrium as a
comparison with the height of the $q=10$ equilibrium peak of 
figure~\ref{fig_eq_sm} shows. The long tails of the peaks
of the $\beta\to\beta_{\max}$ half-cycle decrease rather rapidly
with increasing $n_{\beta}$, so that $S_{\max}(\beta)$ approaches 
its equilibrium value for $\beta>\beta_{\rm peak}^+$.

For $q=2$ (no figure shown) an approach of both peaks to the 
equilibrium peak of $S_{\max}$ is observed, whose height is for $q=2$ 
only about 10\% smaller than the height of the $n_{\beta}=1$ dynamical 
peak. We take this as an indication that in the 
range of our dynamical speeds the phase conversion mechanism is 
always spinodal, independently of the order of the equilibrium
transition. Figure~\ref{fig_1V_sm} makes this point by contrasting
the equilibrium results of figure~\ref{fig_eq_sm} with the 
$n_{\beta}=1$ dynamical results. In the next subsection we analyze 
our structure functions data with respect to this scenario.

The locations of the equilibrium peaks are closer to the 
$\beta^-_{\rm peak}$ values of the dynamical $\beta\to\beta_{\rm min}$ 
heating half-cycles than to the $\beta^+_{\rm peak}$ values of the
dynamical $\beta\to\beta_{\max}$ cooling half-cycles. This is 
particularly clear for $q\ge 4$. Our understanding of this is that the 
relaxation is faster for the heating than for the cooling half-cycle. 
This observation goes hand in hand with the interpretation of the higher 
peaks in figures~\ref{fig_q10_sm}, \ref{fig_q10_smV} and~\ref{fig_1V_sm} 
as being due to faster response times of the systems.

\subsection{Structure Functions} \label{subsec_sf}

During our simulations we recorded the structure function~(\ref{sf}) 
for the following momenta:
\begin{eqnarray} \label{k1}
k_1 &=& (2\pi L^{-1},0)~~{\rm and}~~(0,2\pi L^{-1}) \\
k_2 &=& (2\pi L^{-1},2\pi L^{-1}) \\
k_3 &=& (4\pi L^{-1},0)~~{\rm and}~~(0,4\pi L^{-1}) \\
k_4 &=& (4\pi L^{-1},2\pi L^{-1})~~{\rm and}~~
        (2\pi L^{-1},4\pi L^{-1}) \\
k_5 &=& (4\pi L^{-1},4\pi L^{-1}) \label{k5}
\end{eqnarray} 
The structure functions are averaged over rotationally equivalent
momenta.
In the following we use the notation $S_{k_i}=S_{k_i}(\beta)$, 
$(i=1,\dots,5)$ for the structure function $S(\vec{k},t)$, when
the vector $\vec{k}$ is $k_i$ and the time dependence is dictated 
by $\beta=\beta(t)$. Spinodal decomposition is characterized by
an explosive growth in the low momentum modes, while the high 
momentum modes relax to their equilibrium values.

\subsubsection{Lowest momentum $k_1$}

\begin{figure}[ht] \vspace{-2mm} \begin{center}
\epsfig{figure=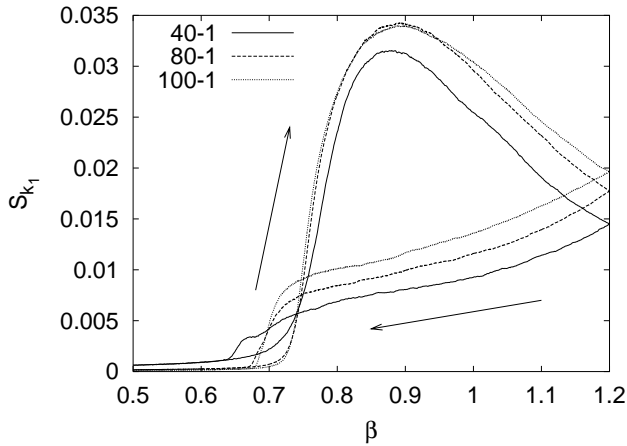,width=\columnwidth} \vspace{-1mm}
\caption{The structure function $S_{k_1}(\beta)$ for the 
10-state Potts model and $n_{\beta}=1$ dynamics.} 
\label{fig_q10_k1} \end{center} \vspace{-3mm} \end{figure}

\begin{figure}[ht] \vspace{-2mm} \begin{center}
\epsfig{figure=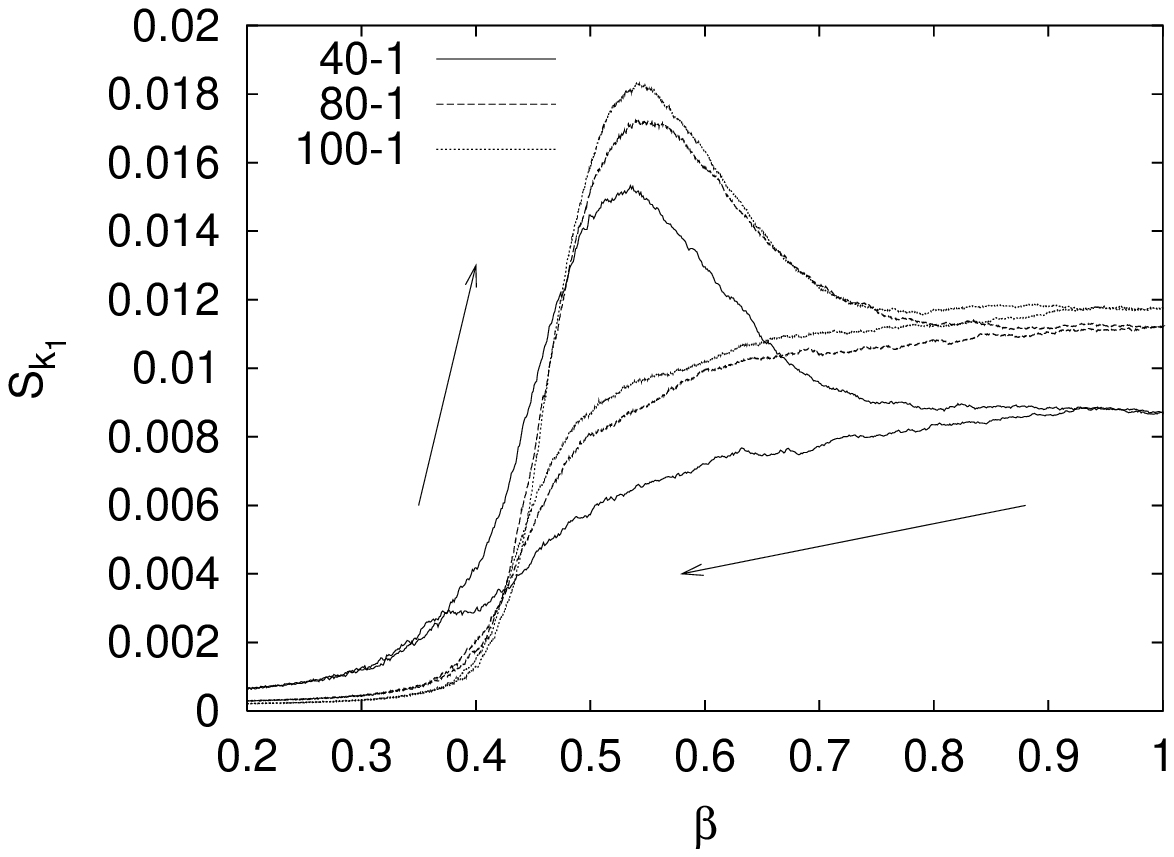,width=\columnwidth} \vspace{-1mm}
\caption{The structure function $S_{k_1}(\beta)$ for the 
2-state Potts model and $n_{\beta}=1$ dynamics.} 
\label{fig_q02_k1} \end{center} \vspace{-3mm} \end{figure}

\begin{figure}[ht] \vspace{-2mm} \begin{center}
\epsfig{figure=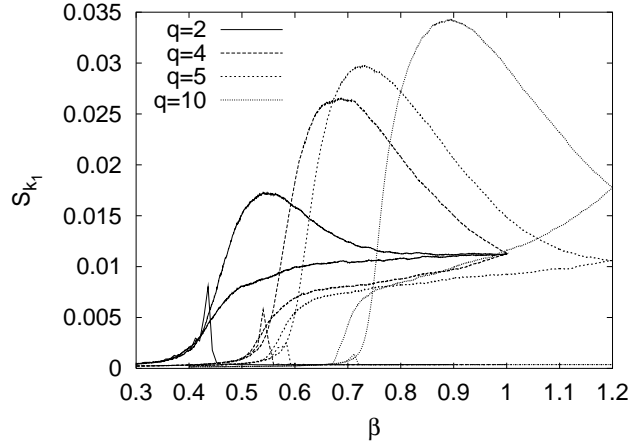,width=\columnwidth} \vspace{-1mm}
\caption{The structure function $S_{k_1}(\beta)$ from $n_{\beta}=1$
dynamical simulations on $80\times 80$ lattices together with their
equilibrium values.} 
\label{fig_sf1k1V} \end{center} \vspace{-3mm} \end{figure}

\begin{figure}[ht] \vspace{-2mm} \begin{center}
\epsfig{figure=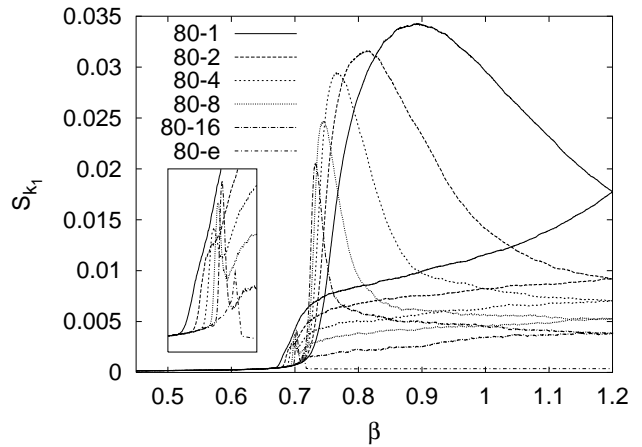,width=\columnwidth} \vspace{-1mm}
\caption{The structure function $S_{k_1}$ for the 10-state Potts
model on $80\times 80$ lattices for dynamical simulations with 
$n_{\beta}$ as indicated by the extensions to the lattice size.
The inlay of the left enlarges the equilibrium peak together with
the heating data. For the inlay $\beta$ is mapped on 
$0.5+2*(\beta-0.66)$ and $S_{k_1}$ on $0.002+4*S_{k_1}$. }
\label{fig_q10_sf1kV} \end{center} \vspace{-3mm} \end{figure}

\begin{figure}[ht] \vspace{-2mm} \begin{center}
\epsfig{figure=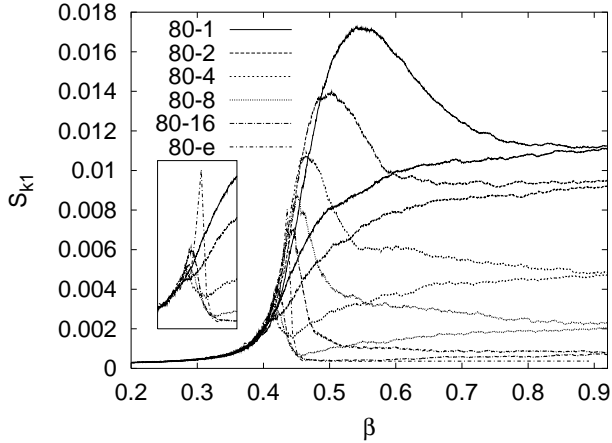,width=\columnwidth} \vspace{-1mm}
\caption{The structure function $S_{k_1}$ for the 2-state Potts
model on $80\times 80$ lattices for dynamical simulations with 
$n_{\beta}$ as indicated by the extensions to the lattice size.
The inlay of the left enlarges the equilibrium peak together with
the heating data. For the inlay $\beta$ is shifted by $-0.13$ and
$S_{k_1}$ by $+0.002$. }
\label{fig_q02_sf1kV} \end{center} \vspace{-3mm} \end{figure}

In figures~\ref{fig_q10_k1} and~\ref{fig_q02_k1} we give the $q=10$
and $q=2$ results 
for $S_{k_1}(\beta)$.  The hysteresis flow is indicated by the arrows.
Figure~\ref{fig_sf1k1V} shows the $q$-dependence of the $n_{\beta}=1$ 
dynamics together with our equilibrium results.
In comparison with the $S_{\max}(\beta)$ figures~\ref{fig_q10_sm}, 
\ref{fig_q02_sm}, \ref{fig_eq_sm} and~\ref{fig_1V_sm}, several 
differences and similarities deserve to be mentioned:

\begin{enumerate}

\item For our $n_{\beta}$ values the $S_{k_1}$ structure functions 
      have a pronounced maximum on the $\beta\to\beta_{\max}$ 
      half-cycle. This is cooling in the spin system language used 
      in this paper and heating (i.e., confinement to deconfinement) 
      in an analogue QCD system. 

\item Different ordinate scales are chosen in figures~\ref{fig_q10_k1}
      and~\ref{fig_q02_k1}, because the magnitudes of the peaks show
      a considerable $q$-dependence. That is exhibited in 
      figure~\ref{fig_sf1k1V}.

\item As in figure~\ref{fig_eq_sm} for $S_{\max}$ the equilibrium 
      peaks of $S_{k_1}$ increase from $q=10$ to $q=2$ (see 
      figure~\ref{fig_sf1k1V}). However, increasing $n_{\beta}$ 
      from~1 to~16, the approach of the $S_{k_1}(\beta)$ function to
      their equilibrium is even for $\beta>\beta_{\rm peak}^+$ rather 
      slow. This is shown for $q=10$ in figure~\ref{fig_q10_sf1kV}.
      To a large extent it holds still for $q=2$, where for 
      $\beta>\beta_{\rm peak}^+$ the equilibrium appears to become 
      approached for $n_{\beta}=16$, as is shown in 
      figure~\ref{fig_q02_sf1kV}. 

\item The inlays of figures~\ref{fig_q10_sf1kV} and~\ref{fig_q02_sf1kV}
      enlarge the equilibrium peaks together with the heating
      ($\beta\to\beta_{\min}$) data. For $q=10$ as well as for $q=2$
      we find that the heating data develop a peak with increasing
      $n_{\beta}$, which may eventually approach the equilibrium
      peak. However, in both cases it appears that the heating peak
      wants first to merge with the cooling peak. For $q=10$ the
      $n_{\beta}=16$ cooling peak is much larger than the equilibrium
      peak and the heating peaks are all the time increasing. Eventually,
      both peaks should start to decrease towards the equilibrium data.
      For $q=2$ the cooling peaks decrease rapidly with increasing 
      $n_{\beta}$ and the $n_{\beta}=16$ cooling peak undershoots 
      the equilibrium peak. Cooling and heating peak move towards 
      merging and should approach the equilibrium peak from below.


\end{enumerate}

The $S_{k_1}$ structure function peaks strongly under cooling and
less under heating. This is presumably related to the fact that the 
spin variables get ordered at low temperatures (like the Polyakov 
loops get ordered at high temperatures), thus allowing for $q$-ality
order-order domains. For our fast dynamics
the $S_{k_1}$ values at $\beta_{\max}$ are so high that the peaks
under heating become overshadowed. This can be made explicit by
first equilibrating the systems at $\beta_{\max}$. Then peaks of 
similar size as the equilibrium peaks appear on the $\beta_{\max}
\to\beta_{\min}$ half-cycle.
For our slower dynamics the $S_{k_1}(\beta)$ structure function
peaks for the $\beta\to\beta_{\min}$ half-cycle become visible 
without equilibrating first at $\beta_{\max}$. For increasing 
$n_{\beta}$ they approach the equilibrium peaks (see the inlays 
of figures~\ref{fig_q10_sf1kV} and~\ref{fig_q02_sf1kV}).
For both half-cycles the approach to equilibrium appears to 
happen only for a really slow dynamics. Very CPU time consuming 
simulations of $n_{\beta}\gg 16$ values would be needed to 
follow this in detail. 

\subsubsection{$k_i,\ i\ge1$ and quenching}

\begin{figure}[ht] \vspace{-2mm} \begin{center}
\epsfig{figure=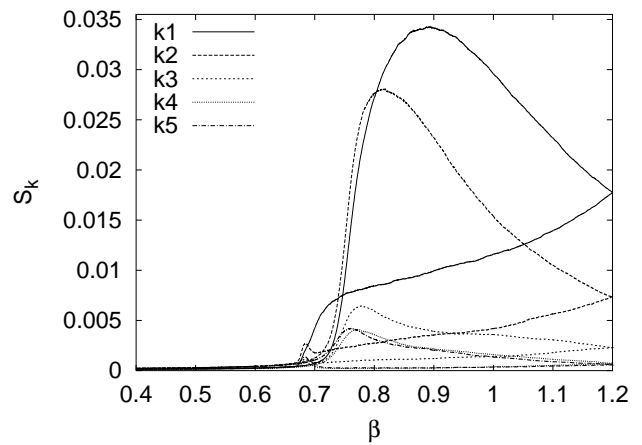,width=\columnwidth} \vspace{-1mm}
\caption{Hysteresis of the $S_k(t)$ structure functions for the 
10-state Potts model on an $80\times 80$ lattice ($n_{\beta}=1$).} 
\label{fig_q10_sfk} \end{center} \vspace{-3mm} \end{figure}

\begin{figure}[ht] \vspace{-2mm} \begin{center}
\epsfig{figure=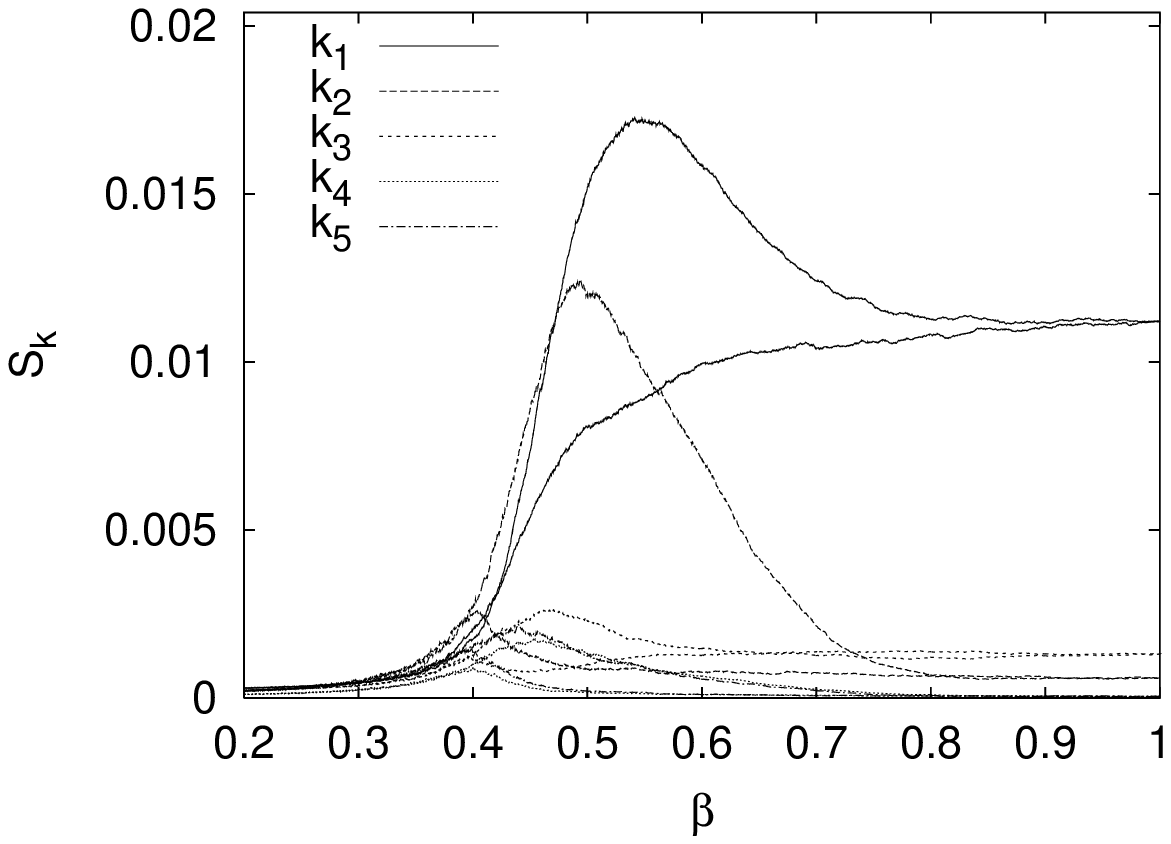,width=\columnwidth} \vspace{-1mm}
\caption{Hysteresis of the $S_k(t)$ structure functions for the 
2-state Potts model on an $80\times 80$ lattice ($n_{\beta}=1$).} 
\label{fig_q02_sfk} \end{center} \vspace{-3mm} \end{figure}

\begin{figure}[ht] \vspace{-2mm} \begin{center}
\epsfig{figure=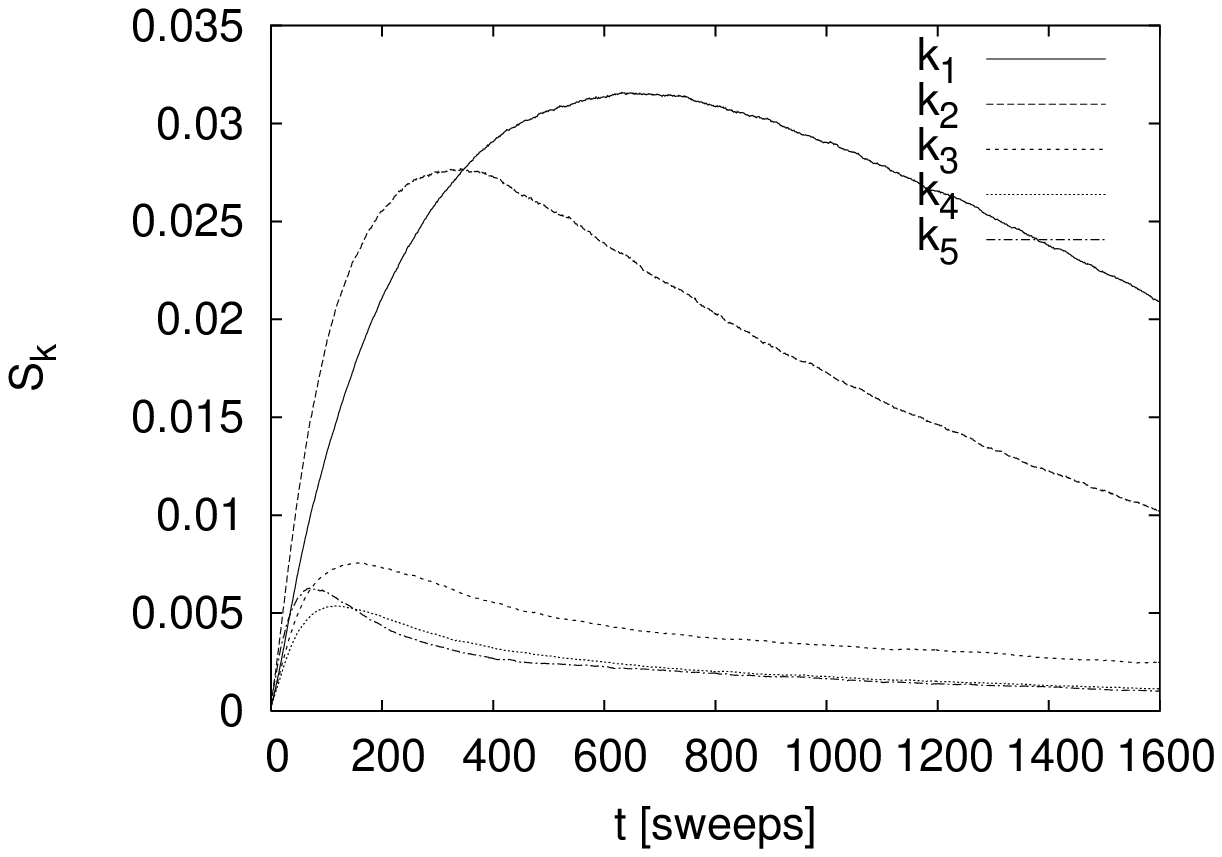,width=\columnwidth} \vspace{-1mm}
\caption{Time evolution of the $S_k(t)$ structure functions for
the 10-state Potts model on an $80\times 80$ lattice
after quenching from $\beta_{\min}=0.4$
to $\beta_{\max}=0.8$.} 
\label{fig_q10_quench} \end{center} \vspace{-3mm} \end{figure}

\begin{figure}[ht] \vspace{-2mm} \begin{center}
\epsfig{figure=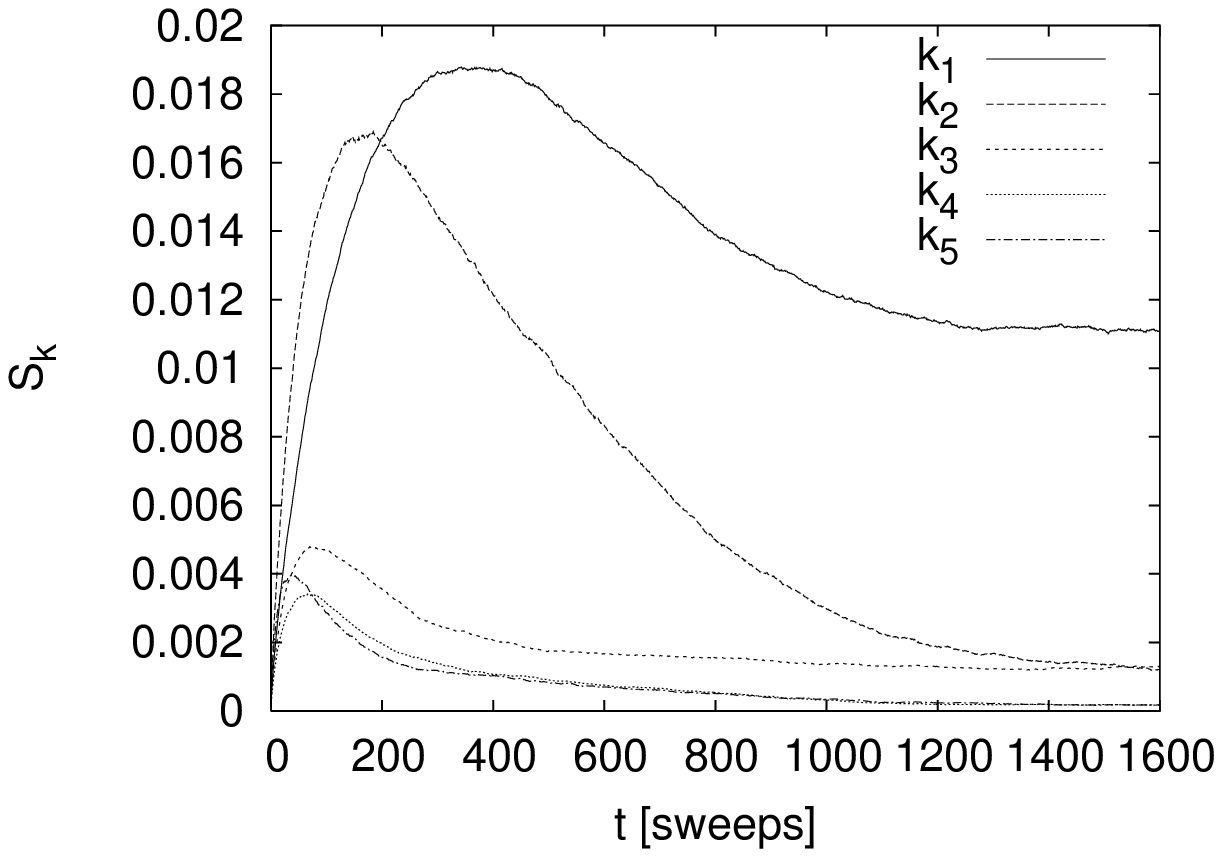,width=\columnwidth} \vspace{-1mm}
\caption{Time evolution of the $S_k(t)$ structure functions for
the 2-state Potts model on an $80\times 80$ lattice
after quenching from $\beta_{\min}=0.2$
to $\beta_{\max}=0.6$.} 
\label{fig_q02_quench} \end{center} \vspace{-3mm} \end{figure}

\begin{figure}[ht] \vspace{-2mm} \begin{center}
\epsfig{figure=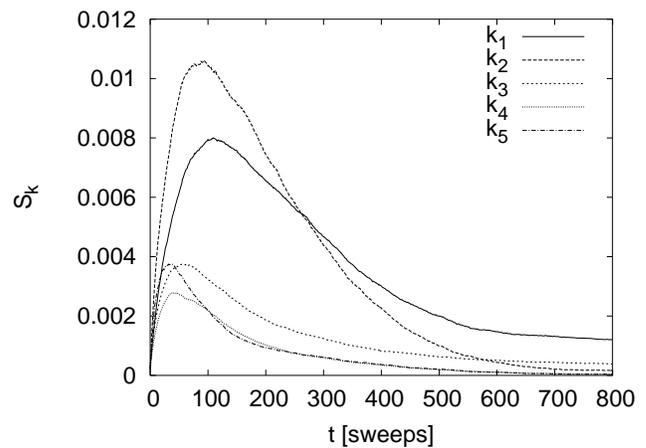,width=\columnwidth} \vspace{-1mm}
\caption{Time evolution of the $S_k(t)$ structure functions for
the 2-state Potts model on an $80\times 80$ lattice
after quenching from $\beta_{\min}=0.2$
to $\beta_{\max}=0.6$ at a magnetic field $h=0.01$.} 
\label{fig_q02_hquench} \end{center} \vspace{-3mm} \end{figure}

Miller and Ogilvie~\cite{MiOg02} investigated the dynamics of
SU(2) gauge theory after quenching from a low to a high physical 
temperature (corresponding to the $\beta_{\min}\to\beta_{\max}$
half-cycle of the spin system). They report a critical value $k_c$,
so that modes grow (do not grow) exponentially for $k<k_c$ ($k>k_c$).

In figures~\ref{fig_q10_sfk} and~\ref{fig_q02_sfk} we show for our
$n_{\beta}=1$ dynamics all structure functions, which we have measured. 
It is notable that
we observe a large gap between the peaks for $S_{k_i}(\beta)$, $i=1,2$
and for $S_{k_j}(\beta)$, $j\ge 3$. We also performed quenching runs. 
In figure~\ref{fig_q10_quench} we show the time evolution on an 
$80\times 80$ lattice after quenching the 10-state Potts model from
$\beta=0.4$ to $\beta=0.8$ and in figure~\ref{fig_q02_quench} we show
the time evolution after quenching the 2-state Potts model from 
$\beta=0.2$ to $\beta=0.6$. As in our hysteresis investigations 
averages of 640 independent repetitions are taken. Again, we
find a large gap between the peaks for $S_{k_i}(\beta)$, $i=1,2$
and for $S_{k_j}(\beta)$, $j\ge 3$. Further it is remarkable that
the height of the peaks in the hysteresis cooling half-cycles and
under quenching are almost identical, while the time scale is 
according to equation~(\ref{delta_beta}) extended to 3200 sweeps 
for the hysteresis curve.

For figure~\ref{fig_q02_hquench} we have changed the second order
phase transition of the 2-state Potts model to a crossover by
adding the term $h\sum_{\vec{r}}\delta_{\sigma(\vec{r},t),\,q_0}$
with a small magnetic field $h$, $h=0.01$, with respect to the spin 
$q_0$ in the Hamiltonian~(\ref{H}). In essence the time evolution
after quenching is similar as without the magnetic field, only
that the magnitude of the $S_{k_1}$ and $S_{k_2}$ peaks decreases, 
maintaining still a clearly visible gap to the peaks of the 
$S_{k_j}$, $j\ge 3$ structure functions. Increasing the magnetic 
field further, to $h=0.1$, the large peaks disappear altogether 
by merging into the small peaks. The signal of a transition is 
possibly lost for such high values of the magnetic field.

The observed peaks suggest a critical $k_i$ value between $i=2$
and $i=3$ for the Potts models. However, the issue is more subtle.
Cahn-Hilliard theory~\cite{CaHi58,Ca68} (model~B for reviews 
see~\cite{ChLu97,La92,Gu83}) predicts an exponential growth of
the low momentum structure function in the initial part of the time 
evolution after quenching (also applies to model~A). Whether such 
an exponential growth is 
found or not was used by Miller and Ogilvie~\cite{MiOg02} to determine
the critical $k_c$ between the low and the high momentum mode.
However, none of our structure functions in figures~\ref{fig_q10_quench}
to~\ref{fig_q02_hquench} shows initial exponential growth. This is
already kind of obvious by looking at the figures, where the shape
of the increasing parts of the curves is always concave and is
quantitatively determined by performing fits. A likely explanation
is that the Cahn-Hilliard theory relies on approximations which
are not justified in our $2d$ models. In $3d$ we find exponential 
growth in the very early stage of the time evolution after quenching 
(incrementing then $\beta$ after every update and fitting the 
time evolution within the first sweep~\cite{BV}). 
Interestingly, our hysteresis curves of figures~\ref{fig_q10_sfk}
and~\ref{fig_q02_sfk}, which rely on the smoother dynamics of 
temperature changes in small steps, show an initial exponential 
growth. 

We find no peaks in the structur factors when we quench from an
ordered initial state into the disordered phase. The reason may
be that the structure factor is defined with respect to the 
order parameter. Under a quench into the ordered phase $q$-ality
order-order domain may emerge, whereas there is only one disordered
phase.

\section{Summary and Conclusions} \label{sec_conclude}

Our energy hysteresis method allows for dynamical estimates of the 
equilibrium transition temperatures for first as well as for second 
order phase transitions. While the precision of these estimates is 
not competitive with those of equilibrium investigations, the 
hysteresis method provides information about dynamically rooted 
deviations of accompanying physical observables from their equilibrium 
values. For second order transitions we find that the dynamics 
generates a latent heat and for a weak first order transition we find 
a `dynamical' latent heat much larger than its equilibrium value, 
whereas for a strong first order transition the dynamical latent heat 
agrees with the equilibrium value (the magnetization allows for a 
similar analysis, which is not reported here). 

In our analysis of $2d$ Potts models, we find spinodal decomposition
to be the dominant feature as soon as we turn on the dynamics. For
instance, the equilibrium (quasi static) phase conversion of first 
order phase transitions is due to nucleation. Even our slowest 
dynamics ($n_{\beta}=16$) changes the phase conversion of the 
investigated weak ($q=5$) and strong ($q=10$) first order transitions 
from nucleation to spinodal. 
For the $q=2$ (weak) second order transition the $n_{\beta}=16$ 
dynamics appears to be already rather close to equilibrium, which 
is formally reached for $n_{\beta}\to\infty$.
These results are mainly based on analyzing the dynamical time 
evolution of Fortuin-Kasteleyn (FK) clusters and structure functions.

\begin{itemize}

\item For FK clusters we find that the largest cluster surface area 
is quite sensitive to dynamical effects and yields for all considered
$q$-values signals in favor of a spinodal decomposition on the cooling 
($\beta\to\beta_{\max}$) and heating ($\beta\to\beta_{\min}$) 
half-cycles of our hysteresis loops. This may be illustrated by
comparing the results of our fast ($n_{\beta}=1$) dynamics of 
figure~\ref{fig_1V_sm} with the equilibrium results of 
figure~\ref{fig_eq_sm}. For the first order transitions the 
dynamics enhances the peak values to take on similar values as 
one finds for the $q=2$ and $q=4$ second order configurations 
near the critical point.

\item For the structure factor our $n_{\beta}=1$ dynamics leads on 
the $\beta_{\min}\to\beta_{\max}$ half-cycle to amplitude maxima, 
which are considerably larger than those from the second order 
equilibrium configurations, see figure~\ref{fig_sf1k1V}. The 
dynamical peaks on the $\beta_{\max}\to\beta_{\min}$ half-cycle 
are comparable to those of from the equilibrium configurations,
which has for figure~\ref{fig_sf1k1V} the consequence that they
are not visible at all, because the systems are still out of 
equilibrium at $\beta_{\max}$ and on their return path. That is 
of potential interest for heavy ion collision, where our 
$\beta_{\min}\to\beta_{\max}$ half cycle corresponds to 
heating for which the dynamics of the experiment is definitely fast.
We have no entirely satisfactory theoretical explanation for the 
observed asymmetry, but think that is is related to the fact
that $q$-ality order-order domain may emerge in the ordered 
(deconfined) phase.

\end{itemize}

Using quenching, we find dynamical signals surviving even after 
the proper (second order) phase transition is converted into a 
crossover. Moving then far away from the transition, the dynamical 
signals fade away too and the issue of crossovers requires further 
investigations.

Our computer programs allow to extend the present study to the 3-state 
Potts model in three dimensions with an external magnetic field 
representing quark effects. In a more remote future, one could 
carry out similar studies for quenched and even full QCD. But none 
of these studies could resolve the problem of a quantitative
 relationship between the Glauber time scale of our Euclidean dynamics
and the time scale of the Minkowskian dynamics in the real world.
In this context it is of interest that Pisarski and Dumitru developed
recently a Polyakov loop model~\cite{DuPi01} which allows for 
simulations in the Minkowskian formulation~\cite{Sc01}. It may be 
possible to address questions similar to those raised in our paper 
within the hyperbolic dynamics of their model. 

Most likely the aim of such studies cannot be to make precise 
quantitative predictions. Instead, one may have to be content 
with illustrating the effect of different speeds of the phase 
conversion on the observable signals qualitatively. If spinodal 
decomposition of Polyakov loops is indeed realized in heavy ion 
collisions, one may observe an enhancement in the production of 
low-energy gluons. 
\medskip

\acknowledgments
BB and AV would like to thank Michael Ogilvie for useful discussions.
This work was in part supported by the US Department of Energy under
contract DE-FG02-97ER41022. The simulations were performed on workstations 
of the FSU HEP group. Testruns were done on workstations of the IUB.


\begin{thebibliography}{19}

\bibitem{MO96} 
S.~Ejiri,
Nucl.\ Phys.\ Proc.\ Suppl.\  {\bf 94} (2001) 19;
F.~Karsch,
Lect.\ Notes Phys.\  {\bf 583} (2002) 209;
H. Meyer-Ortmanns, Rev. Mod. Phys. {\bf 68} (1996) 473.

\bibitem{Br90} Ch. Schmidt, C.R. Allton, S. Ejiri, S.J. Hands, 
O. Kaczmarek, F. Karsch and E. Laermann, Nucl. Phys. B. (Proc. Suppl.)
{\bf 119} (2003) 517;
F. Karsch, AIP Conf. Proc. {\bf 602} (2001) 323 (hep-lat/0109017);
F. Karsch, E. Laermann, and Ch. Schmidt, Phys. Lett. 
B {\bf 520} (2001) 41.  
F.R. Brown, F.P. Butler, H. Chen, N.H. Christ, Z. Dong, W. Schaffer, 
L.I. Unger, and A. Vaccino, Phys. Lett. B {\bf 251} (1990) 181;

\bibitem{MOS96} H. Meyer-Ortmanns and B.-J. Schaefer, Phys. Rev. 
D {\bf 53} (1996) 6586-6601.

\bibitem{MiOg02} T.R. Miller and M.C. Ogilvie, Nucl. Phys. B
(Proc. Suppl.) {\bf 106} (2002) 357; Phys. Lett. B {\bf 488}
(2000) 313.

\bibitem{AKMcL} R. Anishetty, P. Koehler, and L. McLerran, Phys. Rev. 
D {\bf 22} (1980) 2793.

\bibitem{Bj83} J.D. Bjorken, Phys. Rev. D {\bf 27} (1983) 140 and
references given therein.

\bibitem{BKT93} J.D. Bjorken, Int. J. Mod. Phys. A {\bf 7}
(1992) 4189; J.D. Bjorken, K.L. Kowalski and C.C. Taylor, 
hep-ph/9309235.

\bibitem{RaWi93} K. Rajagopal and F. Wilczek, Nucl. Phys. B {\bf 399}
(1993) 395; {\bf 404} (1993) 577; F. Wilczek, Nucl. Phys. A {\bf 566}
(1994) 123.
  
\bibitem{BHMV03} B.A. Berg, U.M. Heller, H. Meyer-Ortmanns, and
A. Velytsky, hep-lat/0308032, to appear in the Proceedings of the
Tsukuba Lattice 2003 Conference, Nucl. Phys. B (Proc. Suppl).
  
\bibitem{Po78} A.M. Polyakov, Phys. Lett. {\bf B72} (1978) 477.
  
\bibitem{SY82} B. Svetitsky and L.G. Yaffe, Nucl. Phys. B
{\bf 210} (1982) 423.
  
\bibitem{Gr83} M. Gross, J. Bartholomew, and D. Hochberg, Phys.
Lett. B {\bf 113} (1983) 218.
  
\bibitem{Og84} M. Ogilvie, 
Phys. Rev. Lett. {\bf 52} (1984) 1369.
  
\bibitem{GoOg85} A. Gocksch and M. Ogilvie, Phys. Rev. D {\bf 31}
(1985) 877.

\bibitem{BU83} T. Banks and A. Ukawa, 
Nucl. Phys. B {\bf 225 [FS9]} (1983) 145.
  
\bibitem{Ba73} R.J. Baxter, J. Phys. {\bf C6} (1973) L445.
  
\bibitem{BJ92} C. Borgs and W. Janke, J. Phys. I France {\bf 2} 
               (1992) 2011, and references given therein.

\bibitem{Wu82} F.Y. Wu, Rev. Mod. Phys. {\bf 54} (1982) 235.
  
\bibitem{Gl63} R.J. Glauber, J. Math. Phys. {\bf 4} (1963) 294;
K. Kawasaki, in {\it Phase Transitions and Critical Phenomena},
C. Domb and M.S. Green (editors), Vol. {\bf 2} (1972) 443.

\bibitem{ChLu97} P.M. Chaikin and T.C. Lubensky, {\it Principles of 
condensed matter physics}, Cambridge University Press, Cambridge, 1997.
  
\bibitem{Rothe} H.J. Rothe, {\it Lattice Gauge Theories -- 
An Introduction}, World Scientific Lecture Notes in Physics Vol.43
(Chapter 17), Singapore, 1992, and references given therein.
  

\bibitem{FK72} C.M. Fortuin and P.W. Kasteleyn, Physica {\bf 57} 
               (1972) 536; A. Coniglia and W. Klein, J. Phys.
               {\bf A13} (1980) 2775.

\bibitem{BV} B.A. Berg, H. Meyer-Ortmanns, and A. Velytsky, 
in preparation.

\bibitem{CJR79} M. Creutz, L. Jacobs and C. Rebbi, Phys. Rev.
Lett. {\bf 42} (1979) 1390.

\bibitem{SW87} R.H. Swendsen and J. Wang, Phys. Rev. Lett. 
               {\bf 58} (1987) 86.

\bibitem{Ca68} J.W. Cahn, Trans. Metall. Soc. AIME {\bf 242} (1968) 166.

\bibitem{Hi61} M. Hillert, Acta Met. {\bf 9} (1961) 525.

\bibitem{La92} J.S. Langer in "Solids far from Equilibrium", C.
Godreche (editor), Cambridge University Press, Cambridge, 1992.

\bibitem{Gu83} J.D.Gunton, M. San Miguel and P.S. Sahni, in "Phase 
Transitions and Critical Phenomena", Vol.8, C. Domb and J.L. Lebowitz 
(editors), Academic Press, London, 1983.


\bibitem{Vel} A. Velytsky, Ph.D. thesis, in preparation.

\bibitem{Ke89} J. Kert\'esz, Physica {\bf A161} (1989) 58. 

\bibitem{StAh94} D. Stauffer and A. Aharony, {\it Introduction to
Percolation Theory}, Taylor \& Francis, London 1994.

\bibitem{Sa01} H. Satz, Comp. Phys. Commun. {\bf 147} (2002) 40, 
and references given therein. 

\bibitem{St71} H.E. Stanley, {\it Introduction to Phase Transitions
and Critical Phenomena}, Clarendon Press, Oxford, 1972, p.98--p.100
and p.204--p.210.

\bibitem{VBH02} A. Velytsky, B.A. Berg and U. Heller, Nucl. Phys. B 
                (Proc. Suppl.) {\bf 119} (2003) 861. 

\bibitem{Fi74} M.E. Fisher, in {\it Nobel Symposium 24}, edited by
B. Lundquist and S. Lundquist (Academic Press, New York, 1974).

\bibitem{BN92} B.A. Berg and T. Neuhaus, Phys. Rev. Lett. {\bf 68} 
               (1992) 9. 

\bibitem{BNB94} A. Billoire, T. Neuhaus and B.A. Berg, Nucl. Phys. 
                B {\bf 396} (1993) 779; B {\bf 413} (1994) 795; 
                T. Neuhaus and J.S. Hager, J. Stat. Phys. {\bf 113}
                (2003) 47.  

\bibitem{BlNi00} H.W.J. Bl\"ote and M.P. Nightingale, 
Phys. Rev. B {\bf 62} (2000) 1089.

\bibitem{CaHi58} J.W. Cahn and J.E. Hilliard, J. Chem. Phys. 
                 {\bf 28} (1958) 258.

\bibitem{DuPi01} R.D. Pisarski, Phys. Rev. D {\bf 62} (2000) 111501(R);
                 A. Dumitru and R. Pisarski, Phys. Lett. B {\bf 504}
                 (2001) 282. 

\bibitem{Sc01} O. Scavenius, A. Dumitru and A.D. Jackson, Phys. Rev.
               Lett. {\bf 87} (2001) 182302 

\end{thebibliography}
\end{document}